\documentclass[11pt,oldfont]{article}
 \usepackage{epsf}
 \usepackage{amsfonts}
 \usepackage{amssymb}
 \usepackage{epsfig}
\usepackage{graphics}
\usepackage{latexsym}

\def\myref#1{(\oldref{#1})}
\let\oldref=\ref
\let\ref=\myref

\def\ub{\mbox{\boldmath$\ub$}}

\def\ub{\mbox{\boldmath$u$}}

\def\B{\begin{equation}}
\def\E{\end{equation}}

 \newfont{\bbbold}{msbm10 scaled \magstep1}

 \newfont{\goth}{eufm10 scaled \magstep1}

 \def\be{\begin{equation}}\def\ee{\end{equation}}
 \def\bea{\begin{eqnarray}}\def\eea{\end{eqnarray}}
 \def\ba{\begin{array}}\def\ea{\end{array}}

 \def\ub{\underline{\phantom{\alpha}}\!\!\!\beta}

 \let\la=\label

 {}

 \def\bd{\begin{document}}
 \def\ed{\end{document}}
 \def\bea{\begin{eqnarray}}\def\barr{\begin{array}}\def\earr{\end{array}}
 \def\eea{\end{eqnarray}}
 \def\ft#1#2{{\textstyle{{\scriptstyle #1}\over {\scriptstyle #2}}}}
 \def\fft#1#2{{#1 \over #2}}
\def\sst#1{{\scriptscriptstyle #1}}
 \def\oneone{\rlap 1\mkern4mu{\rm l}}
 \newcommand{\eq}[1]{(\ref{#1})}
 \def\eqs#1#2{(\ref{#1}-\ref{#2})}
 \def\det{{\rm det\,}}
 \def\tr{{\rm tr}}\def\Tr{{\rm Tr}}
\newlength{\dinwidth}
\newlength{\dinmargin}
\begin{document}

\begin{flushright}  
              \end{flushright}

\thispagestyle{empty}

\vspace*{0.5cm}

\begin{center}{\LARGE {  Field theory insight from the AdS/CFT correspondence
}}

\vskip1cm

Daniel Z.Freedman$^1$ and Pierre Henry-Labord\`ere$^2$ 

\vskip0.2cm

$^1$Department of Mathematics and Center for Theoretical Physics\\
Massachusetts Institute of Technology, Cambridge, MA 01239\\
E-mail: dzf@math.mit.edu  
\vskip0.5cm
$^2$LPT-ENS, 24, rue Lhomond \\
F-75231 Paris cedex 05, France \\
E-mail: phenry@lpt.ens.fr

\vskip1.0cm

\begin{minipage}{12cm}\footnotesize

{\bf ABSTRACT}
\bigskip
A survey of ideas, techniques and results from d=5
supergravity for the conformal and mass-perturbed phases of d=4 
${\cal N}$=4 Super-Yang-Mills theory.

\bigskip
\end{minipage}
\end{center}
\newpage



\section{Introduction}

The $AdS/CFT$ correspondence \cite{mal,magoo,kle,wit} allows one to calculate
quantities of interest in certain d=4 supersymmetric gauge
theories using 5 and 10-dimensional supergravity. Miraculously one gets
information on a strong coupling limit of the gauge theory--
information not otherwise available-- from classical supergravity
in which calculations are feasible. 

The prime example of $AdS/CFT$ is
the duality between ${\cal N}$=4 SYM theory and
D=10 Type IIB supergravity. The field theory has the very special
property that it is ultraviolet finite and thus conformal invariant. 
Many years of elegant work on 2-dimensional $CFT's$ has taught us
that it is useful to consider both the conformal theory and its
deformation by relevant operators which changes the long distance
behavior and generates a renormalization group flow of the couplings.
Analogously, in d=4, one can consider
\\
a) the conformal phase of ${\cal N}$=4 SYM
\\
b) the same theory deformed by adding mass terms to its Lagrangian
\\
c) the Coulomb/Higgs phase.
\\
Conformal symmetry is broken in the last two cases. There is now
considerable evidence that the strong coupling behavior of all 3
phases can be described quantitatively by classical supergravity.
In the conformal phase, the $AdS_5 \times S_5$ ``ground state'' of
D=10 supergravity is relevant, while the phases with $RG$-flow are
described by solitons or domain wall solutions.

The result of work by many theorists over the last few years 
is a quantitative picture of
the strong coupling limit of the theory which is a remarkable
advance on previous knowledge. In this lecture we will survey some
of the ideas, techniques, and results on the conformal and massive
phases of the theory. We attempt to reach non-specialists and
are minimally technical.

\section{The ${\cal N}$=4 SYM theory}

The ${\cal N}$=4 SUSY Yang-Mills theory in d=4 with $SU(N)$ gauge 
group can be obtained by dimensional reduction of d=10 SYM.
The fields consist of a gauge field $A_{\mu}$, 4 Weyl fermions
$\lambda_{a}$ (in $\underline{4}$ of the R-symmetry
group SU(4)) and 6 real scalars $X^{i}$ (in $\underline{6}$  of SU(4)).
Each of these fields can be taken as an $N \times N$ traceless
Hermitian matrix of the adjoint of $SU(N)$. The R-symmetry or
flavor symmetry group will play a more important role in our discussion
than the gauge group.
Explicit calculations have shown that the $\beta$ function of the
theory is zero 
up to three-loop order, and arguments for ultraviolet finiteness
to all orders have been given \cite{mand,mand2,mand3}. Finiteness
implies conformal symmetry. The conformal group of 4-dimensional
Minkowski space is $SO(4,2)$$\sim$$SU(2,2)$. This combines with the R-symmetry
$SO(6)$$\sim$$SU(4)$ and ${\cal N}$=4 SUSY
to give the superalgebra $SU(2,2|4)$ which is the over-arching
invariance of the theory. It contains 4${\cal N}$=16 supercharges
$Q_{\alpha}^a$ associated with Poincar\'e SUSY and 4${\cal N}$=16
additional conformal supercharges.

The observables of the theory are the correlation functions of
gauge invariant operators which are composites of the elementary
fields. These operators are classified in irreducible representations
(irreps) of $SU(2,2|4)$. There are many such operators, but those
of chief interest for $AdS/CFT$ belong to short representations.
The scale dimension $\Delta$ of these is fixed at integer or
1/2-integer values and correlated with the $SU(4)$ irrep. One
basic reference on the irreps of $SU(2,2|4)$ is \cite{Dobrev,Flato}
and there is considerable information in the current literature,
e.g. see \cite{ferr}

Chiral primary operators correspond to lowest weight states of
short irreps.
The most important ones are 
\be
TrX^{k}= TrX^{(i_1}X^{i_2}...X^{i_k)}
\ee
where the parentheses indicate a symmetric traceless tensor
in the $SO(6)$ indices. The rank $k$ chiral primary has
dimension $\Delta$=k, and the Dynkin designation of its
R-symmetry irrep is $(0,k,0)$. The dimensions of these irreps
are 20' for k=2, 50 for k=3, 105 for k=4....
By applying $Q_{\alpha}^{a}$ to these operators, we obtain
descendant operators in the same $SU(2,2|4)$ irrep.
For example, the descendants of $TrX^{2}$ include
the $SU(4)$ flavor currents 
$J_{\mu}^{I}$
and the stress tensor $T_{\mu \nu}$.

The dynamics of ${\cal N}$=4 SYM has been much explored through
the years. We now mention a few aspects of this dynamics which
will be illuminated later, through $AdS/CFT$.
\\
1. Ward identities, anomalies, and ${\cal N}$=1 Seiberg dynamics
 can be combined \cite{Erlich} to show that 2-point functions of
 flavor currents and stress tensor are not renormalized. This
 means that all radiative corrections vanish and the exact
 correlation functions are given by the free-field approximation.
 Using ${\cal N}$=1 conformal superspace \cite{osborn}, this result can be
 extended to 2- and 3-point functions of the lowest chiral
 primary $TrX^2$ and all descendents. Similar results can be
 derived using extended superspace \cite{AFGJ}. It is also known
 that 4-point correlation functions do receive radiative corrections,
 so the theory is not secretly a free theory.
\\
2. The scalar potential of the theory is a positive quartic in $X^i$ of the
form
\be
V(X)=bg_{YM}^2Tr([X^i,X^j])^2
\ee
There is thus a moduli space of supersymmetric minima, ie $V(\langle X
\rangle)$=0,
in which the vacuum expectation values $\langle X^{i} \rangle$ are traceless
diagonal matrices. Generically the preserved gauge symmetry is
$U(1)^{N-1}$, and the particle content at a generic point of
moduli space is $N-1$ massless photons, $N^2-N$ massive gauge
bosons and superpartners. This gives a representation of ${\cal N}$=4
supersymmetry with central charges.
\\
3. We will also be interested in supersymmetric
mass deformations of the theory. It is then convenient to describe
the deformed theory using  ${\cal N}$=1 chiral superfields $\Phi_i
(i={1,2,3})$ whose lowest components are complex superpositions of the
6 $X^i$. The deformed theory has the superpotential
\be
W=g_{YM}Tr\Phi_{3}[\Phi_{1},\Phi_{2}]+{1\over 2}
\sum_{i=1}^{3}m_{i}Tr(\Phi_{i})^{2}
\la{mt}
\ee
The dynamics then turns out to depend very dramatically on the
pattern of the mass parameters $m_i$.
\\
For $m_{3}$$\neq$$0$ and $m_{1}$=$m_{2}$=$0$, the methods of Seiberg
dynamics have been used \cite{leigh} to show that  
conformal symmetry is broken at intermediate scales, but the theory 
flows to a non trivial conformal fixed point in the infrared limit.
\\
For $m_{1}$=$m_{2}$=$m_{3}$=$m$, the supersymmetric vacua obey 
$[\phi^{1},\phi^{2}]$=${m \over g_{YM}}\phi_{3} $ and are in one-to-one
correspondence with representations of $SU(2)$ \cite{HSSW}. The long
distance physics depends on which vacuum is chosen, and gauge
symmetry can be realized in both confined and Higgsed phases.
\\
We have time and space here to discuss only the holographic dual of
the first mass deformation \cite{fgpw}, despite extremely interesting
recent work on the second \cite{pw2,ps,stra}.

\section{D=10 Type IIB Supergravity}
 The bosonic fields of the $D$=10 $IIB$ supergravity consist of a metric $g$, 
a scalar dilaton $\phi$ and axion $C$, two three-forms and a self-dual
five-form $F_{5}$.

The classical equations admit as exact background $AdS_{5} \times S^{5}$
with $F_{5}$=$N vol(S^{5})$. N comes from flux quantization. 
The fields $\phi$ and $C$ are constant and the other fields vanishes. 

The complete Kaluza-Klein mass spectrum on $AdS_{5} \times S^{5}$ was
obtained in \cite{van}. It is organized into super-multiplets whose
component fields are in representations of the $SO(6)$ isometry group
of $S^5$. The lowest multiplet contains the graviton and its
super-partners:

$(g_{\mu \nu},\psi_{\mu}^{\alpha}$ in the
$\underline{4}+\underline{4}^{\ast}$,
 $A_{\mu}$ in the $\underline{15}$, $B_{\mu \nu}$ in the
$\underline{6}_{c}$, $\lambda$ in the
$\underline{4}+\underline{20}+\underline{4}^{\ast}+\underline{20}^{\ast}$,
the scalars $\varphi^{i}$ in the $\underline{1}_{c}+
 \underline{10}+\underline{10}^{\ast}+\underline{20}')$

Each of these fields is the lowest state of a Kaluza-Klein tower of
fields. For example a scalar of the $D$=10 theory can be expanded as
\be
\varphi(z,y) = \sum_{\Delta=1}^{\infty}\varphi_{\Delta}(z)Y^{\Delta}(y)
\ee
Here $z^\mu$ and $y^i$ are coordinates of $AdS_5$ and $S^5$, respectively,
and $Y^{\Delta}(y)$ is a rank $\Delta$ spherical harmonic on $S^5$. The
masses of the 5-dimensional scalars $\varphi_{\Delta}(z)$ are eigenvalues
of the $SO(6)$ Casimir operator, and masses are related to the rank
$\Delta$ by $m^2$=$\Delta(\Delta-4)$.

One can now begin to see the duality between d=4, ${\cal N}$=4 SYM and
the dimensionally reduced D=10 supergravity theory. Type IIB
supergravity has 32 supercharges preserved by the
$AdS_5 \times S^5$ vacuum. The isometry group is $SO(4,2) \times SO(6)$
and there are 4 gravitini, which indicates that the superalgebra
is indeed $SU(2,2|4)$. One can look in more detail and find that
the Kaluza-Klein spectrum of supergravity contains D=5 fields
in exactly the same short representations as those of the chiral
primaries $TrX^k$ of d=4, ${\cal N}$=4 SYM. So there is 1:1
correspondence of fields in the D=5 and d=4 theories with a
perfect match of $SO(6)$ irreps and scale dimensions. For example,
there are scalar fields $\varphi_\Delta(z)$ with $\Delta$=k for
each of operators $TrX^k$. The 15 gauge vectors $A_\mu^I(z)$
of supergravity are dual to the $SO(6)$ flavor currents $J_\mu^I$,
of the field theory and fluctuations of the metric $g_{\mu\nu}(z)$ are 
dual to the stress tensor $T_{\mu \nu}$.

In principle the equations of motion of the D=10 completely
determine the interactions in the $AdS_5 \times S^5$ background.
However, the dimensional reduction process is already very complicated
at the linear level, since the independent (i.e. uncoupled) 
5-dimensional fields are actually
mixtures of those in the 10-dimensional theory \cite{van}. The
nonlinear interactions are even more complicated, and they have been
worked out in only a few sectors of the theory \cite{lmrs}.

On the other hand there is a known complete nonlinear 5-dimensional
supergravity theory, the gauged ${\cal N}$=8 theory of \cite{grw,grw2} which
contains the fields of the graviton multiplet above and is invariant
under the same superalgebra. This D=5 ${\cal N}$=8 theory is believed
to be a consistent truncation of the Type IIB theory on $AdS_5
\times S^5$.
This means that any classical solution of the D=5 theory can be lifted to an
exact solution of the D=10. The Kaluza-Klein ``ansatze'' required to
establish consistent truncation are quite complicated. Thus no complete
proof for the D=10 IIB theory has been given. However there has
been recent progress in finding the lifts of several solutions \cite{pw1,pw2},
and the truncation property has been proven for other 
theories \cite{dewit,nastase}. 

The study of the dynamical implications of the $AdS/CFT$ correspondence
is somewhat limited by the absence of a complete dimensionally
reduced action which includes interactions of all Kaluza-Klein modes.
For perturbative questions (and some others) one can work at the
level of Type IIB on $AdS_5 \times S_5$. One cannot always do this
when exact solutions are needed, and one works instead with the
D=5 ${\cal N}$=8 theory.

\section{ AdS/CFT correspondence}

\subsection{General setting}

Maldacena \cite{mal} conjectured an exact duality between d=4, ${\cal
N}$=4 SYM and Type IIB string theory on $AdS_5 \times S^5$. String theory on
this space-time is not yet well defined, so Maldacena argued further
that duality holds at the level of classical Type $IIB$ supergravity
on  $AdS_5 \times S^5$ provided two conditions hold:
\\
1) the $AdS$ length scale $L$ and string scale $\alpha'$ must satisfy
$L^2$$\equiv$$\alpha'(g^2_{YM}N)^{\frac{1}{2}}>>\alpha'$ so that stringy
corrections to supergravity are small. This requires 
$\lambda$$\equiv$$g^2_{YM}N$ large, i.e. strong coupling in SYM.
\\
2) N$\rightarrow$$\infty$ so that loop corrections in SG or in string
theory can be neglected.
\\
Maldacena's conjectured duality was given dynamical content and predictive
power by  Gubser, Klebanov and
Polyakov \cite{kle} and by Witten \cite{wit}. For our present purposes the
duality means that observables of the field theory such as the correlation 
functions can be calculated from classical supergravity if the two conditions
$N \rightarrow \infty$, $\lambda>>1$ hold.

We will explore the dynamics of the $AdS/CFT$ correspondence in a toy
model of gravity and a scalar field in 5 bulk dimensions. The action is
\be S=\int d^{5}x\sqrt{g}(-{1 \over 4}R+{1 \over 2}{(\partial \phi)}^{2}
-V(\phi))
\ee
where we work in units in which $\kappa_5^2$=$8\pi G_5$=2. We assume that
the potential $V(\phi)$ has one or more critical points $\bar\phi$ at
which  $V(\bar\phi)<0$. The classical equations of motion are
\be
\frac{1}{\sqrt{-g}}\partial_{\mu}(\sqrt{-g}g^{\mu\nu}\partial_{\nu}\phi)+
V'(\phi)=0
\ee
\be
R_{\mu\nu} - \frac{1}{2}
g_{\mu\nu}R=2{\partial_{\mu}\phi\partial_{\nu}\phi
-g_{\mu\nu}[(\partial\phi)^2-2V(\phi)]}
\ee
At each critical point there is a solution with constant $\phi(z)$=$\bar\phi$
and an $AdS$ geometry satisfying
\be
 R_{\mu\nu} - \frac{1}{2} g_{\mu\nu}R = 2 V(\bar\phi) g_{\mu\nu}
\ee
The cosmological constant $\Lambda$ and $AdS_{(d+1)}$ scale are related
 to $V(\bar\phi)$ by  $\Lambda$=$2V(\bar\phi)$=$-\frac{d(d-1)}{L^2}$.  
There are several common coordinate systems in which the $AdS_{(d+1)}$ 
metric can
be presented -- each making different features of the geometry evident.
For the purposes of this lecture the most useful form is
\be
 g=e^{2A(r)}\eta_{ij}dx^{i}dx^{j}-dr^{2} 
\la{metric}
\ee
with the scale factor $e^{2A(r)}$=$e^{{2r \over L}}$ and Minkowski$_d$ metric
$\eta_{ij}$=$(+--\cdots-)$. These coordinates are not global; they cover
only the Poincar\'e patch of the full space-time. The Poincar\'e patch
is natural for the $AdS/CFT$ correspondence because it makes the
relevant symmetries evident (although interesting issues beyond the scope
of this lecture do arise from the non-global property).
The continuous symmetries of this metric include
\\
1) an obvious $d$-dimensional Poincar\'e symmetry group with $\frac{d(d+1)}{2}$
   parameters.
\\
2) the scale tranformation $r$$\rightarrow$$r+a$, $x^i$$\rightarrow$
$e^{{-a \over L}}x^i$.
\\
3) an additional $d$ parameters of special conformal transformations with
usual infinitesimal form $\delta x^i$=$(x)^2c^i -2(c.x)x^i$. Readers
are invited to find the corresponding transformation of $r$.
\\
The complete continuous isometry group is $SO(d,2)$. Thus the isometry
group of $AdS_5$ is the same as the conformal group of Minkowski$_4$.

It is also important to realize that the surface 
$r$$\rightarrow$$\infty$ is technically a boundary. A null geodesic
gets there in finite time, and boundary conditions must be supplied 
to obtain unique solutions of wave equations in the $AdS$ geometry. The
boundary is conformal to d=4 Minkowshi space.
We are interested in stable solutions and classical stability is
determined by the $M^2$ of scalar fluctuations about the background,
$h(z)$$\equiv$$\phi(z)-\bar\phi$. We express the mass in units of the
$AdS$ scale, ie. $M^2$=$V''(\bar\phi)$$\equiv$$m^2/L^2$. An $AdS$ solution
can be stable even for negative $m^2$ provided that the stability
bound \cite{fb,fb2} $m^2$$\geq$$-4$ is satisfied. In more realistic models
there are several scalars $\phi^i$ and masses are eigenvalues $m_{I}^2$
of the hessian matrix $L^2 \frac{\partial^2V}{\partial\phi_i\partial\phi_j}$.
Stability requires $m_{I}^2$$\geq$$-4$.

In realistic models there is a $CFT_4$ associated with each stable
critical point of $V(\phi^i)$ with a map between bulk fields $\phi^{i}$
and field operators ${\cal O}_{\Delta_i}$ whose scale dimension is
$\Delta_i$=$2 +\sqrt{4+m_{i}^2}$. In the $AdS/CFT$ prescription for
correlation functions of the ${\cal O}_{\Delta_i}$, their bulk
duals act as sources in a generating functional. In more detail
the prescription \cite{wit} is:
\\
1. Solve the classical equations of motion for the fluctuations,
\be   
\frac{\delta S}{\delta\phi^i}= \Box \phi^i + \frac{\partial
V}{\partial\phi^i}=0
\ee
subject to boundary conditions as $r \rightarrow \infty$
\be
\phi^i(\vec{x},r) \rightarrow e^{\frac{(\Delta_i-4)r}{L}}
\bar{\phi}^i(\vec x)
\ee
This is a Dirichlet problem modified to accomodate the exponential 
scaling rate found from standard Frobenius analysis of the equation.
\\
2. Substitute the solution into the action to obtain the on-shell
action $S[\bar{\phi}^i(\vec x)]$$\equiv$$S[\phi^i(\vec x,r)]$ and regard this as a functional
of the boundary data.
\\
3. The $CFT_4$ correlators are then given by
\be
\langle {\cal O}_{\Delta}(\vec{x}_1) \cdots  {\cal O}_{\Delta}(\vec{x}_n)
\rangle
=\frac{\delta}{\delta\bar{\phi}(\vec{x}_1)} \cdots
\frac{\delta}{\delta\bar{\phi}(\vec{x}_n)}
e^{iS[\bar{\phi}]}
\ee

The functional prescription can be turned into a precise diagrammatic
algorithm, and some Witten diagrams for 2-, 3-, and 4-point functions
are shown in Fig 1.
\\

\begin{figure}[http]
\centering
\epsfxsize=2in
\epsfysize=2in
\epsffile{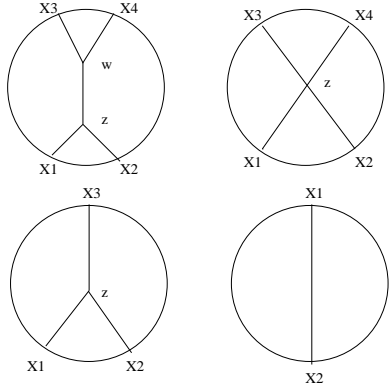}
\caption{ Witten diagrams}
\la{diagr}
\end{figure}

a. a Witten diagram Fig. \ref{diagr} is much like a Feynman diagram. There are boundary
  points $\vec x$ at which the $CFT$ operators are inserted, while bulk
  points $z,w$ are integrated over $AdS_5$
\\
b. There are quite simple bulk-to-boundary propagators
\be
K_\Delta(\vec{z}-\vec{x},z_0)=C_{\Delta}\bigl(\frac{z_0}{(\vec{z}-\vec{x})^2+z_{0}^{2}}
\bigr)^{\Delta}
\ee
They are solutions of the linear wave equation 
$(\Box+\frac{m^2}{L^2})K_\Delta$=$0$ where $\Box$ is the invariant wave operator.
\\
c. For exchange graphs one also needs bulk-to-bulk propagators. They
   are hypergeometric functions of the variable 
$u$=$\frac{(z_0-x_0)^2+(\vec{z}-\vec{x})^2}{2z_0x_0}$.
For scalars, they can be obtained
from the $AdS$ literature of the 1980's.
\\
d. The Feynman rules for bulk vertices come, as expected, from $S_{int}$ 
of the supergravity Lagrangian.
\\
In a long program of work the MIT-UCLA collaboration \cite{fmmr,dmmr}
developed systematic techniques to evaluate the $AdS$ integrals
in these diagrams and extract physical results for field theory.
It was also necessary to derive new bulk-to-bulk propagators for
photons and gravitons \cite{prop}. Witten diagrams automatically
give conformally covariant amplitudes which contain the precise
scaling factors to justify the bulk-boundary relation between
scale dimension and mass given above. This is just kinematics,
but the quantitative agreement found between ${\cal N}$=4 SYM
and IIB SG on $AdS_5 \times S^5$ extends far beyond kinematics
and symmetry. There is space here only for the briefest
summary of the key results.

1. In a tour de force technical calculation, the cubic couplings
of the supergravity scalars dual to the operators $TrX^k$ were obtained
in \cite{lmrs}.  Results were combined with the corresponding $AdS$
integrals in \cite{fmmr} and revealed that the supergravity results
for all 3-point correlators $$\langle TrX^kTrX^lTrX^m \rangle$$ agreed with the
free-field Feynman diagrams in the field theory. This suggested
that this whole family of correlators was not renormalized, at
least in the $N$$\rightarrow$$\infty$, $\lambda>>1$ limit in which
the correspondence is valid. This surprising result was then
investigated at weak coupling in the field theory and it was
shown that order $g^2$ radiative corrections to all 2- and 3-
point correlators of the $TrX^k$
(and order $g^4$ in a few cases) vanish. General arguments
indicating that radiative corrections vanish to all
orders in $g^2$ were developed in \cite{int,hw} and elsewhere.   
Remarkably, the field theory results hold for all N and for
any gauge group \cite{skiba}. Thus an important, unrecognized
property of the field theory was revealed through supegravity.

2. Most 4-point functions obtained from supergravity are not
trivial. One can extract an operator product expansion from
the amplitudes of exchanged graphs. All singular powers agree
with the expected contribution of operator dual to the exchanged
field and its conformal descendents. The schematic form of the
resulting double OPE is
\be
\langle {\cal O}_{\Delta_1}(\vec{x1}) \cdots  {\cal
O}_{\Delta_2}(\vec{x4})
\rangle
=\sum_{\Delta}\frac{C_{\Delta_1\Delta_2\Delta}}{x_{12}^{\Delta_1+\Delta_3-\Delta}}
\frac{C_{\Delta}}{x_{13}^{2\Delta}}
\frac{C_{\Delta_3\Delta_4\Delta}}{x_{34}^{\Delta_3+\Delta_4-\Delta}}
\ee
The supergravity amplitudes contain a single power of
$log(x_{12}^2 x_{34}^2/x_{13}^2 x_{24}^2)$. This can be
interpreted as an anomalous dimension of double trace operators
$:TrX^jTrX^k:$. These operators are not directly in the map between
supergravity fields and the single trace $TrX^k$.
Instead their effects are found in an appropriate
short distance limit of $n$-point functions with $n \geq 4$. The
scale dimensions of most double trace operators are renormalized since
they are typically lowest weight operators of long representations.
The 4-point calculations yield $\Delta_{jk}=j + k +\gamma_{jk}/N^2$.
The $1/N^2$ corrections are strong coupling predictions of $AdS/CFT$.
One curious point is that in a generic bulk supergravity
theory,  exchange graphs can have $log^2$ singularities rather than the
single power that appears in the case of fields and couplings of
the Type IIB theory. Only the single log can be interpreted as
the correction to the scale dimension of an operator.
\\
3. The explicit structure of supergravity amplitudes suggested that
``extremal'' 4-point (and $n\geq4$) correlators are also not
renormalized \cite{freedman}. An extremal 4-point function is 
$$<TrX^{k_1}TrX^{k_2}TrX^{k_3}TrX^{k_4}>$$ in the case $k_1=k_2+k_3+k_4$.
This is another very curious prediction of $AdS/CFT$ which was
subsequently confirmed by order $g^2$ and instanton calculations
\cite{bianchi}. General arguments based on harmonic superspace appeared
soon after \cite{eden}. There are further results of this type
for ``subextremal'' correlators including a prediction from field
theory \cite{eden} later confirmed in supergravity \cite{ehwss}.

The situation may be summarized by saying that an interplay of work
by both supergravity and field theory methods has given much new
information about the conformal phase of the ${\cal N}$=4 SYM theory.
It confirms that the $AdS/CFT$ correspondence has quantitative
predictive power, so we can go ahead and apply it in  non-conformal
settings.

\section{Basics of holographic RG flows}
A conformal field theory can be perturbed by adding a relevant
perturbation
$
A_\Delta \int d^{4}\vec{x}
{\cal O}_{\Delta}(\vec{x})
$ to the action, where the  operators ${\cal O}_{\Delta}$ have scale
dimension $\Delta <4$.
One reason to restrict to relevant deformations is to avoid 
uncontrollable ultraviolet
divergences, but 
relevant deformations are
also more interesting physically since they change the low
energy behavior of the theory. As relevant deformations of
${\cal N}$=4 SYM we shall be interested in the operators
$TrX^2$, $Tr\lambda^2$, and $TrX^3$, that is mass terms and
cubic scalar couplings.  

In this section we will explore the ideas involved in finding
the holographic duals of such perturbed $CFT_4$'s. Eventually
we will be led to the gauged ${\cal N}$=8 D=5 supergravity
which contains all the relevant operators of the parent
Type IIB theory. However it is useful to begin the discussion
in terms of the gravity/scalar toy model of Sec III. We shall
describe the basic ideas \cite{girar,distler} of holograhic $RG$ flows
in this model and then gradually move toward the more realistic
case. 

Symmetry considerations are a useful starting point. In the perturbed
theory $SO(4,2)$ symmetry is valid only at short distances, but the
true spacetime symmetry is reduced to that of the Poincar\'e group in d=4.
Since symmetries of the bulk and boundary theories must match, we
look for classical solutions of the D=5 bulk theory with Poincar\'e
symmetry. The most general Poincar\'e$_4$ metric and scalar configuration with this symmetry
takes the form
\be
 ds^{2}=e^{2A(r)}[(dt)^{2}-(d\vec{x})^{2}]-dr^{2} 
\la{metric}
\ee
\be
\phi=\phi(r)
\ee
(Other equivalent forms are found in the literature and differ
 by change of the radial coordinate $r$.)
If  $\phi$=const and $A(r)$=${r \over L}$, we recover $AdS_{5}$
with enhanced $SO(4,2)$ symmetry, but we will now be more
general.
With this ansatz, the equations of motion become
\be
\phi''(r)+4A'(r)\phi'(r)={ \partial V \over \partial \phi}
\la{wall1}
\ee
\be
A'(r)^{2}=-{1 \over 3}V(\phi)+{1 \over 6}(\phi')^{2}
\la{wall2}
\ee
These coupled non-linear equations are usually difficult to solve.
One thing which can be done quite generally is to linearize about 
the pure $AdS_5$ solution with scale $L$ associated with a 
fixed point $\bar \phi$ near which
\be V(\bar \phi +h)\simeq \frac{1}{L^2}(-3 + {1 \over
2}m^{2}h^2)
\ee
The fixed point is approached by exact solution
as $r$$\rightarrow$$\pm\infty$.
Moreover, letting $m^{2}$=$\Delta(\Delta-4)$, a standard Frobenius
analysis gives the  scaling rates 
\be 
\lim_{r \rightarrow
-\infty}\phi=Ae^{(\Delta-4)\frac{r}{L}}+Be^{-\Delta
\frac{r}{L}}
\ee 

The $AdS$/CFT correspondence gives the following physical interpretation
of the solution approaching the boundary region $r$$\rightarrow$
$+\infty$:
\\ 
1. a generic solution with $A$$\neq$$0$
corresponds to addition of the operator dual to $\phi$ to the $CFT_4$ 
Lagrangian, i.e. $\Delta{\cal L}$=$A{\cal O}_{\Delta}(\vec{x})$.
\\
2. Special solutions with $A$=$0$, $B$$\neq$$0$ correspond to a 
deformation of CFT by the vev 
$\langle{\cal O}_{\Delta}\rangle_{CFT}\sim B$
\cite{kraus}.
\\
Suppose now that $V(\varphi)$ has two fixed points at values
$\varphi_{UV(IR)}$. The non-linear equations will then have a solution 
$\varphi(r)$ which approaches the constants $\varphi_{UV(IR)}$  and 
$A(r)$ which is asymptotic to ${r \over L_{UV(IR)} }$ as $r$ goes to
$\pm\infty$. This is just 
a domain wall which interpolates between the boundary region of
one  $AdS_{5}$ geometry with
scale $L_{UV}$ and the deep interior region of another $AdS_{5}$ geometry with
scale $L_{IR}$.
The scalar field profile will typically have dominant scale behavior
$A_{UV}$$\neq$$0$ as $r$$\rightarrow$$\infty$. In order to approach $\bar
\varphi_{IR}$ as 
$r$$\rightarrow$$-\infty$, it must behave as $h(r)$$\rightarrow$
$e^{(\Delta_{IR}-4)r}$ with $\Delta_{IR}>4$.
This is the gravity dual of a field theory which flows from a $CFT_{UV}$ 
perturbed by the
relevant operator ${\cal O}_{\Delta}$, toward a different $CFT_{IR}$ at long
distance along an RG trajectory corresponding to perturbation of
the latter theory by an
irrelevant operator.

\subsection{A c-theorem}
When a $CFT_4$ is coupled to a curved external metric $g_{ij}(\vec x)$,
the expected invariance under the Weyl transformation
$\delta g_{ij}(\vec x) =2\delta\sigma(\vec x) g_{ij}(\vec x)$
is broken due to ultraviolet divergences. An anomalous contribution
to the vacuum expectation value $\langle T^i_i \rangle$ of the form
\be
\langle T_{i}^{i} \rangle ={c \over 16\pi^{2}}W_{ijkl}^{2}-{a \over
16\pi^{2}}\tilde{R}_{ijkl}^{2}
\la{ano}
\ee
is then generated. 
The first term is the square of the Weyl tensor and the second is the
topological Euler density. The Weyl anomaly coefficients $c$ and $a$, also
known as conformal central charges, are important data of a $CFT_4$.
They can also be obtained from the long and short distance limits of
correlation functions of the stress tensor. In ${\cal N}$=4 SYM theory
the anomaly coefficients obey non-renormalization theorems
\cite{Erlich} and can
be calculated using only the free-field content of the theory. The
result is $c$=$a$=$\frac{N^2-1}{4}$.

For $CFT_2$ there is the fundamental
Zamolodchikov $c$-theorem which uses the 2-point function of the
stress tensor to construct a positive monotonic function interpolating
 between the central charges of the $UV$ and
$IR$ $CFT$'s in any RG flow. Thus one always has $c_{UV}>c_{IR}$
for the unique central charge in 2-dimensions. There is no
generally accepted proof of a similar theorem in 4-dimensions \cite{Cardy},
but there is a great deal of evidence from model examples
\cite{Cappelli,Erlich} that the Euler anomaly coefficient does satisfy
$a_{UV}>a_{IR}$ (while the same inequality for $c$ is violated
in many cases).

In a very elegant paper \cite{Hen}, Henningson and Skenderis have
shown how to obtain the conformal anomaly holographically. Formally
the on-shell action which is the generating function
for field theory correlators is invariant under 5-dimensional
diffeomorphisms. However, the diffeomorphism group contains a subgroup
which induces Weyl transformations of the boundary metric $g_{ij}(\vec x)$, 
so one
really should expect an anomaly. Indeed the on-shell action is
actually divergent.  It must be
cutoff at a large finite value of $r$ and counter terms
added to cancel the divergence. The counter terms are local functions
of the boundary metric whose variation generates
the trace anomaly in the form \ref{ano}. The specific counter terms
that appear guarantee that $c$=$a$ in any field theory which has a
holographic dual in this framework. Further the holographic value
agrees with the field theory result for ${\cal N}$=4 SYM. 

It is very curious that for holographic RG flows, one can prove
the $c$-theorem \cite{girar} that remains elusive in field theory. 
To do this
one defines the (dimensionless) function $$a(r) = \frac{1}{G_5
A'(r)^3}=\frac{N^2}{4L^3A'(r)^3}$$ 
whose boundary limit agrees with the Henningson-Skenderis calculation
of the central charge. It is easy to manipulate the domain wall
equations \ref{wall1}, \ref{wall2} to show that $A''(r)$=$-2\phi'(r)^{2}/3<0$ so that
the function $a(r)$ decreases monotonically as one moves from the
boundary into the interior of the space-time. This result is
actually independent of the specific dynamics of the bulk theory,
because it simply follows from the Einstein equations for a metric
of the domain wall form that $A''$=$-8\frac{\pi G_5}{3}(T^{t}_t-T^{r}_r) < 0$.
The inequality $(T^{t}_t-T^{r}_r) > 0$ is one of the standard positive
energy conditions in general relativity. A direct consequence of
monotonicity is the c-theorem in the form $a_{UV}>a_{IR}$.

\begin{figure}[http]
\centering
\epsfxsize=2in
\epsfysize=2in
\epsffile{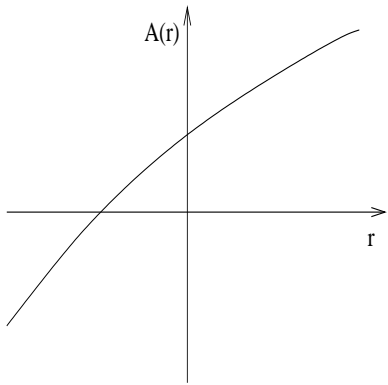}
\la{vv}
\caption{Profile of the scale factor $A(r)$}
\end{figure}

The result $A''(r) <0$ also implies that the profile of the
scale factor $A(r)$ is always concave downward, as shown in
Fig. \ref{vv}. Since $A'(\pm \infty)$=$\frac{1}{L_{(UV,IR)}}$ and 
$V(\bar \phi)$=$-\frac{3}{L^2}$ at a critical point, it is also the
case that the $IR$ endpoint of the flow occurs at a deeper
critical point of the potential than the $UV$ endpoint.

\begin{figure}[http]
\centering
\epsfxsize=1.5in
\epsfysize=1.5in
\epsffile{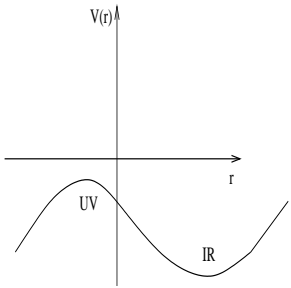}
\caption{Potential $V(\phi)$}
\la{f}
\end{figure}

Let us suppose that the potential $V(\phi)$ is as shown in Fig. \ref{f}.
The critical point labelled $UV$ is a maximum because we are
speaking of a relevant deformation, while that labelled $IR$ is a
minimum since it describes an irrelevant direction in the $CFT_{IR}.$
We have been discussing the classical solution which interpolates
between these two extrema. 

There is another classical solution which departs from the $UV$
critical point and moves to the left as r decreases into the
interior. If there is no other critical point in this direction
the geometry eventually develops a curvature singularity, sometimes
at finite and sometimes at infinite geodesic distance from an
interior point. Such singularities are a major problem for the
$AdS/CFT$ correspondence, since infinite curvature means that
that the supergravity approximation to string theory is invalid.
In some cases stringy mechanisms to resolve the singularity have
been studied \cite{jpp,ps}. In other cases one attempts
a physical interpretation in spite of the singularity by
requiring that fluctuations about the background are regular
at the singularity. Gubser \cite{gubser} has formulated a thermodynamic
criterion for acceptable singularities. 

Let us now discuss an interesting approach to the problem of
solutions of the equations \ref{wall1}, \ref{wall2} for a domain wall 
background.
The approach arose from the study of supersymmetric domain
walls in both D=4 \cite{cvetic} and D=5 \cite{fgpw}, but
was shown to apply more broadly \cite{wfgk,sken}.
Let us define an auxiliary function of $\varphi$, the
superpotential $W(\varphi)$ as follows: \be
{1 \over 8}({dW \over
d\varphi})^2-{1 \over 3}(W)^2=V(\varphi) 
\la{sup}
\ee
and suppose that one can solve this as an ODE for $W(\varphi)$ given
$V(\varphi)$.
Then consider the following set of ODE's 
\be 
{d\varphi \over dr}={1 \over 2}W'(\varphi)
\la{flow1}
\ee
\be
A'(r)=-{1 \over 3}W(\varphi(r))
\la{flow2}
\ee
which can be easily solved by successive quadrature. 
One can then easily show that the solution $\varphi(r)$, $A(r)$ is also a
solution of the original 2nd order gravity-scalar equations
\ref{wall1}, \ref{wall2}.
This turns out to be equivalent \cite{dbvv} to applying Hamilton-Jacobi
theory to the original dynamical system with $W(\varphi)$ as the
Hamilton-Jacobi
function. Unfortunately, it is usually hard to find analytic solutions
for $W(\varphi)$, especially in realistic case of several $\varphi^{i}$ when 
the H-J equation becomes a PDE. It is worth pointing out that the first order
scalar equation then generalizes to the gradient flow equations
\be
{d\varphi^{i} \over dr}={1 \over 2}{dW(\varphi^{j}) \over d\varphi^i}
\la{flow3}
\ee

\subsection{SUSY Flows}
This leads to the miracle of SUSY RG flows: the transformation rules
of a d=5 supergravity theory generate the superpotential $W(\varphi)$ in 
an indirect and rather surprising way. Given 
$W(\varphi)$, one can either solve the first order flow 
equations exactly or gain insight or accurate numerical solutions from 
the well developed theory of gradient flow equations. There is another
important reason to restrict to $SUSY$ flows. It is only for
supersymmetric deformations of ${\cal N}$=4 SYM, that methods of
Seiberg dynamics give sufficient control of the $IR$ behavior of
the boundary field theory. 

To see how all this happens, consider a d=5 supergravity theory with a 
five-bein $e^{m}_{\mu}$, several scalars $\varphi^{i}$, other bosonic
fields
$A_{\mu}^{I}$, $B_{\mu \nu}^{\alpha}$ and fermionic fields $\chi^{A}$
and $\psi_{\mu}^{a}$. The other fields are not ``turned on'' in a
domain wall background since that would violate Lorentz invariance,
but these fields are important as fluctuations dual to operators in
the field theory.

The fermionic transformation rules have the form:
\be
\delta\psi_{\mu}^{a}=D_{\mu}\epsilon^{a}-{1 \over
6}W_{b}^{a}\gamma_{\mu}\epsilon^{b}
\la{Kil1}
\ee
\be
\delta\chi^{A}=(\gamma^{\mu}P_{\mu}(\varphi)^{A}_{a}-Q^{A}_{a}(\varphi))\epsilon^{a}
\la{Kil2}
\ee
The matrices $W^{a}_{b}$, $P_{\mu a}^A$ and $Q^{A}_{a}$ are
functions of scalars $\varphi^{i}$ which are part of the specification of
the classical supergravity theory.
Killing spinors $\epsilon^{a}(\vec{x},r)$ are spinor configurations 
which satisfy
$\delta\psi_{\mu}^{a}$=0 and $\delta\chi^{A}$=0. When they exist they
contain $4n$ arbitrary real parameters, and the background is
said to preserve  $4n$ supercharges. This symmetry is then
matched in the boundary field theory which posesses ${\cal N}$=n 4-dimensional
Poincar\'e supersymmetry.

Let us examine the $\delta\psi_{\mu}^{a}$=$0$ condition and outline how
it leads to the flow equation \ref{flow1}, \ref{flow2}. In the
obvious diagonal local frame for 
the domain wall metric \ref{metric}, the spin connection is

\[\omega_{\mu a b}\sigma^{a b}=\left\{
\begin{array}{ll}
-A'(r)\gamma_{j}\gamma_{5} & \mbox{$\mu=j$} \\
0 & \mbox{$\mu=5$}
\end{array}
\right. \]

The condition \ref{Kil1} can be written in detail as
\be
\delta\psi_{j}^{a}=\partial_{j}\epsilon-\frac{1}{2}A'(r)\gamma_{j}\gamma_{5}
\epsilon^{a}-\frac{1}{6}W^{a}_{b}\gamma_{j}\epsilon^{b}=0
\ee
We can drop the first term because the Killing spinor must be translation
invariant. What remains is a purely  algebraic condition, and we can
see that the flow equation \ref{flow2} for the scale factor directly emerges
with superpotential $W(\phi)$ identified as one of the eigenvalues of
the tensor $W^a_b$. In detail one actually has a symplectic eigenvalue
problem, with 4 generically distinct $W$'s as solutions. Each of these
is a candidate superpotential. One must then
examine the 48 conditions  $\delta \chi^{A}=0$ to see if SUSY is
supported on one of the eigenspaces, and this leads to the gradient
flow equation \ref{flow3}. Success is not guaranteed and generically occurs
on one of the four (symplectic) eigenspaces, giving ${\cal N}$=1 SUSY.
Extended  ${\cal N}>1$ SUSY requires further degeneracy of the
eigenvalues. This process can be implemented reasonably efficiently
using Mathematica programs (if you have the right collaborators).

We now outline the specific application in \cite{fgpw} to the first of 
the several known $RG$ flows found within the gauged D=5 ${\cal N}$=8
supergravity theory. 
\\
1. There are 42 scalars in the bulk theory, and no human being or computer
   can handle the eigenvalue problem analytically in such a large setting.
   One uses generalized symmetry arguments to truncate to possible flows
   involving a small number of fields. In the specific case we describe 
   the final truncation involved 2 scalars and preserved an
   $SU(2)\times U(1)$ flavor symmetry.
\\
2. The Killing spinor conditions were found to be satisfied on a
   subspace of 2 canonically normalized scalars called $\phi_1$ and 
   $\phi_3$ (with $\rho$=$exp(\phi_3/ \sqrt6)$). The superpotential
   found was
\be
W(\phi_{3},\phi_{1})=\frac{1}{\rho^{2}}[ch(2\phi_{1})(\rho^{6}-2)-3\rho^{6}-2]
\ee
\\

\begin{figure}[http]
\centering
\epsfxsize=2in
\epsfysize=2in
\epsffile{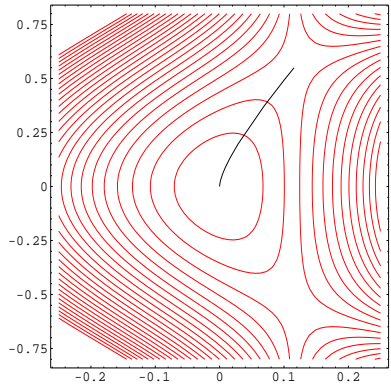}
\caption{Contour plot of $W(\phi_3,\phi_1)$}
\la{plot}
\end{figure}

3. A contour plot of $W(\phi_3,\phi_1)$ is shown in Fig. \ref{plot} and
   reveals three critical points.
\\ 
  a. there is a local maximum at the origin which has the full $SU(4)$
    symmetry and anomalies of the unperturbed ${\cal N}=4$ SYM which
    is the $UV$ endpoint of the flow.
\\
  b. another pair of saddle points at $\phi_{3}= ln(2)/ \sqrt{6}$, 
    $\phi_{1}= \pm ln(3) / 2$. There is a $Z_2$ symmetry under reflection of
    $\phi_1$, so we can restrict to the upper half-plane.
\\
4. The fields $\phi_3,\phi_1$ are $SU(2)\times U(1)$ singlets so all
   gradient flow trajectories have this symmetry. We are most 
   interested in the critical trajectory which terminates at the
   saddle point b. According to our general discussion the
   associated geometry approaches the deep interior of an $AdS_5$
   with scale $L_{IR}$. This determines an anomaly coefficient
   $a_{IR}$=$c_{IR}$= $27N^2/128$. As pointed out in \cite{klm}, the
   anomaly and symmetries, including ${\cal N}$=1 SUSY, exactly match
   those of the ${\cal N}$=4 SYM deformed by the Leigh-Strassler
   mass term $\Delta {\cal L} = m Tr\Phi_{3}^2 / 2$. This is evidence
   that the critical
   trajectory describes the holographic dual of the $RG$ flow in
   that theory.
\\

5. The gradient flow equations for $W(\phi_3,\phi_1)$ are easy to
   write down, but they cannot be solved analytically. An accurate
   numerical determination of the critical trajectory is not 
   difficult, but an analytic solution would be very 
   useful\footnote{ One of the authors (DZF) offers a gourmet dinner
   to the first person to find analytic solutions. See http://www.ClayMath.org/
   for problems with more substantial rewards.}
\\
6. One can check the holographic description of the dynamics of the field
   theory by computing the mass eigenvalues of all fields in the
   theory, namely all fields in the graviton multiplet listed in
section 3,
   at the $IR$ critical point. Scale dimensions are then assigned
   using the formula $\Delta = 2 +\sqrt{4+m^2}$ for scalars and
   its generalizations to other spins.
\\
7. The next step is to assemble component fields into multiplets of
  the $SU(2,2|1)$ superalgebra. There are several short multiplets,
  including chiral multiplets for which the formula $\Delta$=$3R/2$
  relating $U(1)_R$ charge to scale dimension is a helpful constraint.
  Results on the 8 short multiplets agree perfectly with the field
  theory deBernard.Julia@lpt.ens.frscription. This constitutes a non-trivial confirmation
  of the supergravity description of the dynamics of the field theory.
  There are additional results on 3 previously unrocognized semi-short
  mulitplets, headed by the 3 supercurrents of the ${\cal N}$=4 theory
  which are broken by the ${\cal N}$=1 perturbation, and one long
  multiplet. Their scale dimensions are non-perturbative predictions
  of the holographic description.
\\
8. There is an infinite number of other gradient flow trajectories
  emerging from the ${\cal N}=4$ critical point. Generically the
  associated geometries, obtained from the flow equation for $A(r)$
  have curvature singularities. There is a reasonable physical
  interpretation of the trajectories with $\phi_1(r)$=0 as the
  supergravity dual of a Coulomb branch deformation of ${\cal N}$=4 SYM
  \cite{fgpw2,brand}. Since there is only one scalar active in the flow,
  the domain wall profile can be found analytically \footnote{no dinner
for this one}. Some 2-point correlation functions can be calculated
and exhibit a mass gap and other features which are not fully
understood from the viewpoint of the strongly coupled field theory.
\\ 
There is much more to be said about the active subject of holographic  
$RG$ flows, and many interesting papers, including \cite{gppz2,ps,dbvv}
and more, that deserve study by interested theorists.

\bigskip

{\bf Acknowledgments.}
\\

We thank K. Skenderis for a useful discussion. D.Z Freedman also 
acknowledges the hospitality and support of the Centre Emile Borel in
Paris during autumn 2000 and the research support of the National
Science Foundation grant PHY-97-22072

\end{document}